\documentclass[12pt]{article}

\usepackage{amsmath}
\usepackage{amssymb}
\usepackage{amsfonts}
\usepackage{lscape}

\begin{document}

\fontsize{12}{6mm}\selectfont
\setlength{\baselineskip}{2em}

$~$\\[.35in]
\newcommand{\dss}{\displaystyle}
\newcommand{\raro}{\rightarrow}
\newcommand{\be}{\begin{equation}}

\def\sech{\mbox{\rm sech}}
\thispagestyle{empty}

\begin{center}
{\Large\bf Determination of the}   \\    [2mm]
{\Large\bf Electromagnetic Lagrangian } \\    [2mm]
{\Large\bf from a System of Poisson Brackets}   \\  
\end{center}

\vspace{1cm}
\begin{center}
{\bf Paul Bracken}                        \\
{\bf Department of Mathematics,} \\
{\bf University of Texas,} \\
{\bf Edinburg, TX}  \\
{\bf 78541-2999}  \\
{bracken@panam.edu}
\end{center}

\vspace{3cm}
\begin{center}
{\bf Abstract}
\end{center}

The Lagrangian and Hamiltonian formulations of electromagnetism
are reviewed and the Maxwell equations are obtained from the Hamiltonian
for a system of many electric charges. It is shown that three of the equations
which were obtained from the Hamiltonian, namely the Lorentz force law
and two Maxwell equations, can be obtained as well from a 
set of postulated Poisson brackets. It is shown how the results
derived from these brackets can be used to reconstruct the
original Lagrangian for the theory aided by some reasoning
based on physical concepts.

PACS: 45.20Jj, 45.10Db, 02.30.Zz, 03.50.De

\newpage

\begin{center}
{\bf 1. Introduction.}
\end{center}

The Maxwell equations provide a type of mathematical summary of
several fundamental laws of electromagnetism which
originally had their origin in experimental observations. The 
subsequent development of a non-Abelian extension of the Maxwell
theory to the non-Abelian Yang-Mills form and the
diverse applications of Yang-Mills to the subatomic realm of high energy
physics has probably given further impetus to the study of
Maxwell's theory.

A very novel derivation of a pair of the four Maxwell
equations was originally introduced by Feynman, but
the exact details of his argument remained unpublished until
some of the essential arguments were presented by Dyson {\bf [1]}.
It has also been shown that this procedure can be generalized as 
well to the case of the dynamics of particles which possess
other internal degrees of freedom {\bf [2]}. The idea at the
heart of these derivations is to postulate a fundamental set
of Poisson brackets between the fundamental variables of 
the system. The basic defining relations of the bracket are then
applied to the original collection of brackets as well as other
operations, such as differentiation, to generate further new
relations and connections between the variables of the problem {\bf [3]}.
This process is capable of generating some of the Maxwell equations,
as we will show.

In this paper, we begin by developing the Hamiltonian formulation
of the Maxwell theory, and derive the Maxwell equations in this context.
It is shown that the Hamiltonian formalism of the classical system,
like the Lagrangian formalism on which it is based, is also
invariant under gauge transformations {\bf [4-5]}. Different Hamiltonians can 
be written so that they all have the same form, often referred to as
minimal electromagnetic coupling. Next, the development of the
Maxwell equations from a minimal set of defining brackets
involving the variables and the basic algebraic properties of the
Poisson bracket is reviewed {\bf [3]}. To this end, a fundamental set of
brackets is postulated at the outset in a natural way, and
subsequently the algebraic properties of the bracket as well 
defined analytic operations are applied to obtain the basic Lorentz force 
law as well as a pair of the Maxwell equations. Based on these 
initial results, some of the ideas of the inverse problem of the
calculus of variations are applied {\bf [6]}. 
To emphasize, the procedure relies on defining a
basic set of brackets and using fundamental properties
of the bracket to generate new relations,
such as the Leibnitz rule and Jacobi identity,
regardless of the underlying definition of the bracket
It is shown from these
results and with the help of some additional physical motivation at the end that the
full Lagrangian for the theory can be reconstructed.
In the sense that a Lagrangian
theory can be formulated out of a set of elementary results, the
complete set of Maxwell's equations can be obtained.
The results that are obtained from the Lagrangian by means of the 
Euler-Lagrange equations, can then be used to define a Hamiltonian
for the theory to complete the construction {\bf [7-8]}.

It may be asked why this approach is adopted. 
There are several approaches already that begin
this kind of development with commutators {\bf [1]}.
Here we show it is possible to proceed entirely in the
classical domain. Of course, the Maxwell equations
exited before and independently of quantum mechanics,
and nonetheless they are fundamental in
generating quantum theories of electromagnetism.
The same type of analysis can be carried out on
classical theories of gravity and it may prove
possible to adopt some of the ideas here to
proceed to quantum theories of gravity.

\begin{center}
{\bf 2. Lagrangian and Hamiltonian Formalism.}
\end{center}

The system which is of interest here consists of
a collection of nonrelativistic particles which interact with an
external electromagnetic field. The Lagrangian for the system
is sufficient to be used with the principle of least action to 
generate the equations of motion. Moreover, a Lagrangian
is required to construct a Hamiltonian for the system.

A system of nonrelativistic particles, each having a charge
$e_{\alpha}$, mass $m_{\alpha}$ and a displacement vector 
${\bf r}_{\alpha} (t)$ at time $t$ for $\alpha=1,2 , \cdots, N$
in an electric field ${\bf E} ( {\bf x}, t)$ and
magnetic field ${\bf B} ( {\bf x}, t)$ can be described by
the Lagrangian $L$, which in Lorentz-Heaviside units takes
the form
$$
L = \frac{1}{2} \int d^{3} x \, ({\bf E}^{2}  ( {\bf x},t) -
{\bf B}^{2} ( {\bf x}, t)) + \frac{1}{2} \sum_{\alpha=1}^{N} 
m_{\alpha} \dot{\bf r}^{2}_{\alpha}  + \frac{1}{c} \int
d^{3} x \, ( {\bf J} \cdot {\bf A} - c \rho A_{0}) - U.
\eqno(2.1)
$$
For the sake of generality, an arbitrary static external
potential energy $U$ can be included in the Lagrangian as well,
but it is not essential for what follows. The charge density
$\rho ({\bf x},t)$ of the system (2.1) is defined by
$$
\rho ( {\bf x}, t) = \sum_{\alpha=1}^{N} e_{\alpha} \delta 
( {\bf x} - \bf {r}_{\alpha} (t)),
\eqno(2.2)
$$
and the current density ${\bf J} ({\bf x},t)$ associated with the motion
of the particle is given by
$$
{\bf J} ( {\bf x}, t) = \sum_{\alpha =1}^{N}
e_{\alpha} \dot{\bf r}_{\alpha} \delta ({\bf x} - {\bf r}_{\alpha} (t)).
\eqno(2.3)
$$
Of course, the charge density and current density satisfy the
equation of continuity.

The total electromagnetic field is characterized by the
vector potential ${\bf A}$ and a scalar potential $A_{0}$ in
an arbitrary gauge. The total electric and magnetic fields
 are related to the potentials in the following way
$$
{\bf E} ({\bf r}, t) = - {\bf \nabla} A_{0} - \frac{1}{c}
\frac{\partial {\bf A}}{\partial t},
\qquad
{\bf B} = {\bf \nabla} \times {\bf A},
\eqno(2.4)
$$
where ${\bf \nabla}$ is a vector operator and
${\bf \nabla}_{\alpha}$ is the vector operator corresponding to
particle $\alpha$.
The equations of motion can be obtained from Hamilton's
principle of least action by varying the action $S$
$$
S = \int_{t_{1}}^{t_{2}} dt \, L.
$$
The resulting Euler-Lagrange equations take the general form
$$
\frac{d}{dt} \, \frac{\partial L}{\partial \dot{Q}_{j}}
= \frac{\partial L}{\partial Q_{j}} - \sum_{i} \partial_{i}
\frac{\partial L}{\partial (\partial_{i} Q_{j})},
\eqno(2.5)
$$
where the $Q_{i}$ pertain to any of the physical variables
on which $L$ depends. For example, taking $Q$ to be $A_0$,
variation of the action $S$
with respect to $A_{0}$ gives Gauss's law. This can be obtained
as well from the Euler-Lagrange system (2.5)
by identifying $Q_{i}$ with $A_{0}$ and determining the 
derivatives in (2.5).

The Hamiltonian formalism for the total system is manifestly
gauge invariant, and can be determined by first calculating
the canonical momenta. The canonical momentum conjugate to
the coordinate ${\bf r}_{\alpha}$ is given using (2.5) as
$$
{\bf p}_{\alpha} = \frac{\partial L}{\partial \dot{\bf r}_{\alpha}}
= m \dot{\bf r}_{\alpha} + \frac{e_{\alpha}}{c} {\bf A} ( {\bf r}_{\alpha},t),
\eqno(2.6)
$$
The canonical momentum conjugate to the field ${\bf A}$ is
$$
{\bf \Pi} = \frac{\partial L}{\partial \dot{\bf A}} = - \frac{1}{c} {\bf E}.
\eqno(2.7)
$$
Since the Lagrangian is independent of the quantity $\dot{A}_{0}$,
the canonical momentum conjugate to $A_{0}$ is $\Pi_{0} =0$. The 
Hamiltonian for the total system is defined by
$$
H = \int d^{3} x \, ( {\bf \Pi } \cdot \dot{\bf A} + \Pi_{0} \dot{A}_{0})
+ \sum_{\alpha =1}^{N}
{\bf p}_{\alpha} \cdot \dot{\bf r}_{\alpha} - L.
\eqno(2.8)
$$
Substituting the canonical momenta ${\bf \Pi}, \Pi_{0}$
and $\dot{\bf r}_{\alpha}$ in terms of ${\bf p}_{\alpha}$ from (2.6),
as well as the Lagrangian $L$ given in (2.1), we obtain that
$$
H = \sum_{\alpha=1}^{N} \frac{1}{m_{\alpha}} {\bf p}_{\alpha} \cdot 
( {\bf p}_{\alpha} - \frac{e_{\alpha}}{c} {\bf A})
- \frac{1}{2} \int d^{3} x \, ( {\bf E}^2 - {\bf B}^2)
- \frac{1}{c} \int d^3 x \, ( {\bf E} \cdot \dot{\bf A})
$$
$$
- \frac{1}{2} \sum_{\alpha=1}^{N} \frac{1}{m_{\alpha}}
({\bf p}_{\alpha} - \frac{e_{\alpha}}{c} {\bf A})^2 +U - \frac{1}{c}
\int d^3 x \, ( {\bf J} \cdot {\bf A} - c \rho A_{0}).
\eqno(2.9)
$$
Replacing $\dot{\bf A}$ in the third term of (2.8) using (2.4),
we can write
$$
H = \frac{1}{2} \int d^3 x \, ( {\bf E}^2 + {\bf B}^2)
+ \frac{1}{2} \sum_{\alpha=1}^{N} \frac{1}{m_{\alpha}}
({\bf p}_{\alpha} - \frac{e_{\alpha}}{c} {\bf A})^2 + U
+ \int d^3 x \, {\bf E} \cdot {\bf \nabla} A_{0}
$$
$$
+ \sum_{\alpha=1}^{N} \frac{e_{\alpha}}{m_{\alpha} c} {\bf A} \cdot
( {\bf p}_{\alpha} - \frac{e_{\alpha}}{c} {\bf A})
- \frac{1}{c} \int d^3 x \, ( {\bf J} \cdot {\bf A} - c \rho A_{0}).
\eqno(2.10)
$$
Grouping the last three terms in (2.10) together, this equation
can be rewritten using the expressions for $\rho$ and ${\bf J}$ 
given in (2.2) and (2.3)
$$
H = \frac{1}{2} \int d^3 x \, ( {\bf E}^2 ({\bf x},t) + {\bf B}^2 ({\bf x},t) )
+ \sum_{\alpha=1}^{N}
\frac{1}{2 m_{\alpha}} ( {\bf p}_{\alpha} - \frac{e_{\alpha}}{c}
{\bf A} ( {\bf r}_{\alpha},t))^2 + U + \int d^3 x \,
( \rho - {\bf \nabla } \cdot {\bf E}) A_{0}.
\eqno(2.11)
$$
A total derivative term has been dropped to write the Hamiltonian
$H$ in the form (2.11).

From the Hamiltonian (2.11), Hamilton's equations can be developed.
The first of Hamilton's equations for particle $\alpha$ is found
by differentiating $H$ with respect to ${\bf p}_{\alpha}$,
$$
\dot{\bf r}_{\alpha} = \frac{\partial H}{\partial {\bf p}_{\alpha}}
= \frac{1}{m_{\alpha}} ({\bf p}_{\alpha} - \frac{e_{\alpha}}{c} {\bf A}).
\eqno(2.12)
$$
Using (2.12) and the vector identity ${\bf \nabla}_{\alpha}
( \dot{\bf r}_{\alpha} \cdot \dot{\bf r}_{\alpha}) = 2
(\dot{\bf r}_{\alpha} \cdot {\bf \nabla}_{\alpha}) \dot{\bf r}_{\alpha}
+ 2 \dot{\bf r}_{\alpha} \times ({\bf \nabla}_{\alpha} \times \dot{\bf r}_{\alpha})$,
we obtain the second of Hamilton's equations,
$$
\dot{\bf p}_{\alpha} = - \frac{\partial H}{\partial {\bf r}_{\alpha}}
= - \frac{e_{\alpha}}{c} \dot{\bf r}_{\alpha} \cdot {\bf \nabla}_{\alpha}
{\bf A} ( {\bf r}_{\alpha},t) - \frac{e_{\alpha}}{c} \dot{\bf r}_{\alpha}
\times ( {\bf \nabla}_{\alpha} \times {\bf A} ({\bf r}_{\alpha},t))
+ \frac{\partial U}{\partial {\bf r}_{\alpha}} + e_{\alpha}
{\bf \nabla}_{\alpha} A_{0} ( {\bf r}_{\alpha},t).
\eqno(2.13)
$$
Solving for ${\bf p}_{\alpha}$ in (2.12) and differentiating this with
respect to time, we find
$$
\dot{\bf p}_{\alpha} = m_{\alpha} \ddot{\bf r}_{\alpha} + \frac{e_{\alpha}}{c}
( \dot{\bf A}_{\alpha} + \dot{\bf r}_{\alpha} \cdot \frac{\partial {\bf A}_{\alpha}}
{\partial {\bf r}_{\alpha}}).
\eqno(2.14)
$$
Substituting (2.14) into (2.13), and simplifying with equation (2.4) yields
$$
m \ddot{\bf r}_{\alpha} = e_{\alpha} {\bf E} ( {\bf r}_{\alpha},t)
+ \frac{e_{\alpha}}{c} \dot{\bf r}_{\alpha} \times {\bf B}
( {\bf r}_{\alpha}, t) - \frac{\partial U}{\partial {\bf r}_{\alpha}}.
\eqno(2.15)
$$
This is Newton's second law in terms of the Lorentz force.

It remains to complete this process for the field variables
and the procedure is the same as for a system with a finite
number of degrees of freedom. The functional derivative of $H$
is calculated
with respect to ${\bf \Pi}$ defined by (2.7), and this gives the first of
Hamilton's equations 
$$
\dot{\bf A} = \frac{\not\partial H}{\not\partial {\bf \Pi}} =
- c ({\bf E} + {\bf \nabla} A_{0}),
\eqno(2.16)
$$
where the time derivative of ${\bf A}$ is a partial derivative.

The second of Hamilton's equations is given by
$$
\dot{\bf \Pi} = - \frac{\not\partial H}{\not\partial {\bf A}}
= - {\bf \nabla} \times {\bf B} + \frac{1}{c} {\bf J},
\eqno(2.17)
$$
which is the Amp\'ere-Maxwell law,
$$
{\bf \nabla} \times {\bf B} = \frac{1}{c} {\bf J} + \frac{1}{c}
\frac{\partial {\bf E}}{\partial t}.
\eqno(2.18)
$$
The equation for $\Pi_{0}$ is given by
$$
\dot{\Pi}_{0} = - \frac{\not{\partial} H}{\not\partial A_{0}}
= - \rho + {\bf \nabla} \cdot {\bf E}.
\eqno(2.19)
$$
The scalar potential in ${\bf E}$ given in (2.4) is not varied since
${\bf E}$ is proportional to ${\bf \Pi}$, hence independent.
Now, since $\Pi_{0} =0$, equation (2.19) implies Gauss's law,
$$
{\bf \nabla} \cdot {\bf E} = \rho.
\eqno(2.20)
$$
Finally, the equation for $\dot{A}_{0}$ is found by differentiating
$H$ with respect to $\Pi_{0}$. Since $\Pi_{0} = 0$, this equation
is meaningless and does not provide an equation. 
This procedure has in fact generated all four of
Maxwell's equations, although two have not yet been explicitly written down. 
To obtain the remaining two equations, first
take the curl of (2.16) and use the identity ${\bf \nabla}
\times {\bf \nabla} A_{0} = {\bf 0}$ and (2.4), 
$$
{\bf \nabla} \times {\bf E} = - {\bf \nabla} \times
{\bf \nabla} A_{0} - \frac{1}{c} \frac{\partial}{\partial t}
{\bf \nabla} \times {\bf A} = - \frac{1}{c}
\frac{\partial {\bf B}}{\partial t}.
\eqno(2.21)
$$
This is Faraday's law.
The condition that magnetic monopoles do not exist follows
by taking the divergence of ${\bf B}$ given in (2.4),
$$
{\bf \nabla} \cdot {\bf B} = 0.
\eqno(2.22)
$$
To summarize, the Maxwell's equations are given by (2.18),
(2.20), (2.21) and (2.22).

\begin{center}
{\bf 3. Maxwell Equations and Poisson Brackets.}
\end{center}

The algebra of classical observables on the manifold $M$
will be denoted by ${\cal F}$. A Poisson structure on a manifold
$M$ is a skew-symmetric bilinear map which is denoted
$ \{ , \} : {\cal F} \times {\cal F} \rightarrow  {\cal F}$
such that $(i)$ $( {\cal F}, \{ , \})$ satisfies the Jacobi
identity $\{ G, \{ H, K \} \} + \{ H, \{ K , G \} \} +
\{ K, \{ G, H \} \} =0$, $(ii)$ the map $X_{G} = \{, G \}$
is a derivation of the associative algebra ${\cal F } (M)$ on
$M$ {\bf [9]}. It satisfies the Leibnitz rule $\{ G, H \,  K \} = H \{ G , K \}
+ \{ G, H \} K$. A manifold $M$, which is endowed with a
Poisson bracket on ${\cal F} (M)$, is called a Poisson manifold.
These basic algebraic properties are used  on their own
without reference to a specific form for the bracket
to develop the stated results.

Let the local coordinate variables on the manifold be written
in the form $(w^{a}) = ( q^{i}, v^{i})$, where $i=1,2,3$.
Indices will be raised and lowered in a trivial way using
$\delta_{ij}$, and repeated indices are summed over.
Here, the $q^i$ may be interpreted as position coordinates
and $v^{i}$ represent velocity components. The Poisson 
brackets are postulated in the following way {\bf [2-3]},
$$
\{ q_{i}, q_{j} \} = 0,
\qquad
m \{ q_{i}, v_{j} \} = \delta_{ij}.
\eqno(3.1)
$$
Any function $H \in {\cal F}$ will define a dynamical system
on $M$ by the equation
$$
\frac{d G}{d t} = \{ G, H \}.
\eqno(3.2)
$$
Let us postulate an $H \in {\cal F}$ such that equations of
motion can be obtained from (3.2) as follows,
$$
\dot{q}^{i} = \{ q^{i}, H \} = v^{i},
\qquad
m \dot{v}^{i} = m \{ v^{i}, H \} = F^{i}.
\eqno(3.3)
$$
Differentiating the second bracket in (3.1) with respect to
time generates the equation
$$
\{ \dot{q}_{i}, v_{j} \} + \{ q_{i} , \dot{v}_{j} \} = 0.
\eqno(3.4)
$$
Multiplying both sides by $m$ and substituting the equations
of motion, there results the expression
$$
m \{ \dot{q}_{i}, \dot{q}_{j} \} + \{ q_{i}, F_{j} \} = 0.
\eqno(3.5)
$$
Since the bracket is bilinear, this equation can be put
into the form
$$
\{ \{ q_{i}, F_{j} \}, q_{k} \} + m \{ \{ \dot{q}_{i},
\dot{q}_{j} \}, q_{k} \} = 0.
\eqno(3.6)
$$
Substituting $\dot{q}_{i}$, $\dot{q}_{j}$ and $q_{k}$ into
the Jacobi identity, we have
$$
\{ \{ \dot{q}_{i}, \dot{q}_{j} \} , q_{k} \}
+ \{ \{ \dot{q}_{j}, q_{k} \} , \dot{q}_{i} \}
+ \{ \{ q_{k}, \dot{q}_{i} \} , \dot{q}_{j} \} = 0.
\eqno(3.7)
$$
The bracket $\{ \dot{q}_{j}, q_{k} \}$ is proportional to
$\delta_{jk}$ by (3.1), so (3.7) reduces to the constraint
$$
\{ \{ \dot{q}_{i}, \dot{q}_{j} \} , q_{k} \} = 0.
\eqno(3.8)
$$
Substituting (3.8) into (3.6), we obtain that
$$ 
\{ q_{k}, \{ q_{i}, F_{j} \} \} = 0.
\eqno(3.9)
$$
The tensor $\{ q_{i}, F_{j} \}$ is therefore antisymmetric due
to the bracket property. This can be expressed in its dual form
by the relation
$$
\{ q_{i}, F_{j} \} = - \frac{e}{mc} \epsilon_{ijk} B_{k}( {\bf r},t).
\eqno(3.10)
$$
Substituting (3.10) into (3.9), a bracket which contains
$q_{l}$ and $B_{k}$ can be obtained
$$
\{ q_{l}, B_{k} \} = 0.
\eqno(3.11)
$$
The postulated relations (3.1) imply that the vector ${\bf B}$
depends only on the position and time of the particle. Equations (3.11)
and (3.1) imply that $F_{i}$ is at most linear in the
velocities, and so we may write
$$
F_{i} ({\bf r},t) = e E_{i} ({\bf r},t) +
\frac{e}{c} \epsilon_{ijk} v^{j} B^{k} ( {\bf r},t).
\eqno(3.12)
$$
This is the Lorentz force law and serves to define the electric
field. Using the property of bilinearity and the derivation property, we have
$$
\{ q_{i}, e E_{j} + \frac{e}{c} \epsilon_{jak} v_{a} B_{k} \}
= e \{ q_{i}, E_{j} \} + \frac{e}{c} \epsilon_{jak} \{ q_{i}, v_{a} B_{k} \}
$$
$$
= e \{ q_{i}, E_{j} \} + \frac{e}{c} \epsilon_{jak} \{ q_{i}, v_{a} \} B_{k}
+ \frac{e}{c} \epsilon_{jak} v_{a} \{ q_{i}, B_{k} \}
\eqno(3.13)
$$
$$
= e \{ q_{i}, E_{j} \} + \frac{e}{mc} \epsilon_{jik} B_{k}
= e \{ q_{i}, E_{j} \} - \frac{e}{mc} \epsilon_{ijk} B_{k}.
$$
Since the result of (3.13) is the left-hand side of (3.10), 
it follows that
$$
\{ q_{i}, E_{j} \} = 0.
\eqno(3.14)
$$
This bracket implies that the vector ${\bf E}$,  as in the case of ${\bf B}$,
depends only on the time and position coordinates. Equations (3.5) and (3.10)
can be combined and will lead to a new equation for $B_{k}$ 
in terms of the bracket
$$
B^{s} = \frac{ m^2 c}{ 2 e}  \epsilon^{sij} \{ \dot{q}_{i}, \dot{q}_{j} \}.
\eqno(3.15)
$$
Applying the Jacobi identity to the variables $\dot{q}_{l}$, $\dot{q}_{j}$
and $\dot{q}_{k}$ and then contracting with $\epsilon^{ljk}$, there results
$$
\epsilon^{ljk} \{ \dot{q}_{l}, \{ \dot{q}_{j}, \dot{q}_{k} \} \} = 0.
$$
Replacing the bracket $\{ \dot{q}_{j}, \dot{q}_{k} \}$ in this using (3.15)
generates a new bracket involving ${\bf B}$,
$$
\{ \dot{q}_{l}, \epsilon^{ljk} \{ \dot{q}_{j}, \dot{q}_{k} \} \} 
= \frac{2e}{m^2 c} \{ \dot{q}_{l}, B_{l} \} = 0.
$$
This implies that $\{\dot{q}_{l}, B_{l} \} =0$, and therefore
$$
{\bf \nabla} \cdot {\bf B} = 0.
\eqno(3.16)
$$
To obtain a second equation, let us begin with the equation for 
$B_{s}$, given in (3.15). Differentiating both sides with respect to t,
we have
$$
\frac{\partial B_{s}}{\partial t} + \frac{\partial B_{s}}{\partial q_{j}}
\dot{q}^{j} = \frac{m^2 c}{2 e} \epsilon^{sij} \{ \ddot{q}_{i}, \dot{q}_{j} \}
+ \frac{m^2 c}{2 e} \epsilon^{sij} \{ \dot{q}_{i}, \ddot{q}_{j} \}
= \frac{m^2 c}{e} \{ \ddot{q}_{i}, \dot{q} \}.
\eqno(3.17)
$$
Substituting (3.12) into the right hand side of (3.17), dividing out the 
common factor of $e$,
and using property $(ii)$ of the Poisson bracket, there results
$$
m \epsilon^{sij} \{ E_{i} + \frac{1}{c} \epsilon_{ial} \dot{q}_{a} B_{l}, \dot{q}_{l} \}
= m \epsilon^{sij} \{ E_{i}, \dot{q}_{j} \} + \frac{m}{c} \{ B^{s}, \dot{q}_{j} \} \dot{q}_{j}
$$
$$
+ \frac{m}{c} \{ \dot{q}_{j}, \dot{q}_{j} \} B^{s} - \frac{m}{c} \{ \dot{q}^{s},
\dot{q}_{j} \} B_{j} - \frac{m}{c} \dot{q}^{s} \{ B_{j}, \dot{q}_{j} \}.
\eqno(3.18)
$$
The second to last term on the right hand side of (3.18) is zero
by symmetry, and upon substitution of the equation $\{ \dot{q}_{l}, B_{l} \} = 0$
into (3.18), this expression reduces to the following
$$
\frac{\partial B_{s}}{c \: \partial t} + \frac{\partial B_{s}}{c \: \partial q_{j}}
\dot{q}^{j} = - \epsilon_{sji} \frac{\partial E_{i}}{\partial q_{j}}
+ \frac{1}{c} \frac{\partial B_{s}}{\partial q_{j}} \dot{q}_{j}.
$$
Simplifying this, the following Maxwell equation is obtained in the 
usual form
$$
\frac{1}{c} \frac{\partial {\bf B}}{\partial t} + {\bf \nabla} \times {\bf E} =
{\bf 0}.
\eqno(3.19)
$$

\begin{center}
{\bf 4. Reconstruction of the Lagrangian.}
\end{center}

The fundamental Poisson brackets have been assumed to have
the structure given in (3.1), in particular, if we combine
(3.1) and (3.3), the basic relation is of the form
$$
m \{ q^{i}, \dot{q}^{j} \} = \delta^{ij}.
\eqno(4.1)
$$
Consider the classical equations of motion with all
masses set to unity of the form
$$
\ddot{q}_{i} = f_i ( q , \dot{q}, t).
$$
A nonsingular matrix $W_{ij}$ and a function $L (q, \dot{q},t)$
are sought such that
$$
W_{is} (\ddot{q}^{s} - F^{s} )
= \frac{d}{dt} \frac{\partial L}{\partial \dot{q}^{i}} -
\frac{\partial L}{\partial q^{i}}.
\eqno(4.2)
$$
The conditions for the existence of $W_{ij}$ and $L$ are called
the Helmholtz conditions {\bf [10]}. If a Lagrangian $L$ exists, then 
$W_{ij}$ is given by
$$
W_{ij} = \frac{\partial^{2} L}{\partial \dot{q}^{i}
\partial \dot{q}^{j}}.
\eqno(4.3)
$$
From (4.1), we take $W_{ij}$ to be proportional to $\delta_{ij}$.
For the Hessian (4.3) to be invertible, the Lagrangian must obey
$$
\frac{\partial^{2} L}{\partial \dot{q}^{i} \partial \dot{q}^{j}} = 
m \delta_{ij}.
\eqno(4.4)
$$
Integrating (4.4), it can be seen that if $L$ exists, it must
have the form
$$
L = \frac{1}{2} m \dot{q}^{i} \dot{q}^{j} \delta_{ij} + \frac{e}{c} 
\dot{q}^{i} \, A_{i} - e A_{0}  + {\cal C}
= \frac{1}{2} m \dot{\bf r}^{2} + \frac{e}{c} \dot{\bf r} \cdot {\bf A}
- e A_{0} + {\cal C} ({\bf E}, {\bf B}, {\bf A}, A_{0}).
\eqno(4.5)
$$
where the quantity ${\cal C}$ can be regarded as a constant of
integration which will be a functional of the fields which were
introduced in producing the solution (3.12) and equations (3.16), (3.19).
Such a term would represent an energy contribution associated with these
fields, and its exact structure can be determined next on physical grounds. The existence
of $L$, however, does follow from the Helmholtz equations,
as will be seen. If some physical ideas are now introduced and
applied, the Lagrangian (4.5) can be generalized to the structure
given in (2.1) by using (4.5) and writing
$$
L = \sum_{\alpha=1}^{N} \frac{1}{2} m_{\alpha} \dot{\bf r}_{\alpha}^{2} +
\frac{1}{c} \int d^3 x \, ( {\bf J} \cdot {\bf A} - c \rho A_{0})
+ {\cal C} ({\bf E}, {\bf B}, {\bf A}, A_{0}),
\eqno(4.6)
$$
where $\rho$ and ${\bf J}$ have been defined in (2.2) and (2.3).

We say that the force ${\bf F} (t, q^{j}, \dot{q}^{j}, \ddot{q}^{j})$
is potential if Lagrange's equations of motion
$$
\frac{\partial T}{\partial q^{i}} - \frac{d}{dt} \frac{\partial T}{\partial \dot{q}^{i}}
= F_{i},
$$
are variational, that is, if there exists a Lagrange function $L$ such that
$$
\frac{\partial T}{\partial q^{i}} - \frac{d}{dt} \frac{\partial T}{\partial \dot{q}^{i}} -
F_{i} = \frac{\partial L}{\partial q^{i}} - \frac{d}{dt} \frac{\partial L}{\partial \dot{q}^{i}}.
$$
The necessary and sufficient conditions for a force
$F^{i} (t, q^{j}, \dot{q}^{j})$ to be potential can be written
$$
\begin{array}{c}
\dss \frac{\partial F_{i}}{\partial \dot{q}^{j}} +
\dss \frac{\partial F_{j}}{\partial \dot{q}^{i}} = 0,   \\
    \\
\dss \frac{\partial F_{i}}{\partial q^{j}} -
\dss \frac{\partial F_{j}}{\partial q^{i}} +
\dss \frac{d}{dt} \frac{\partial F_{j}}{\partial \dot{q}^{i}} = 0.
\end{array}
\eqno(4.7)
$$
The first of these can be differentiated with respect to
$\dot{q}^{j}$ to give
$$
\frac{\partial^{2} F_{i}}{\partial \dot{q}^{j} \partial \dot{q}^{k}}
= - \frac{\partial^{2} F_{k}}{\partial \dot{q}^{i} \partial \dot{q}^{j}}
= - \frac{\partial^{2} F_{j}}{\partial \dot{q}^{i} \partial \dot{q}^{k}}
= - \frac{\partial^{2} F_{i}}{\partial \dot{q}^{j} \partial \dot{q}^{k}},
$$
therefore,
$$
\frac{\partial^{2} F_{i}}{\partial \dot{q}^{j} \partial \dot{q}^{k}} = 0.
$$
This equation can now be integrated to give the result $F_{i} = a_{ij} \dot{q}^{j}
+ b_{i}$, where $a_{ij}$ and $b_{i}$ are functions of $(t, q^{k})$.
Substituting $F_{i}$ into the pair (4.7), we obtain a set of 
three conditions on the $a_{ij}$ and $b_{i}$ as follows
$$
a_{ij} = - a_{ji},
$$
$$
\frac{\partial a_{is}}{\partial q^{j}} +
\frac{\partial a_{sj}}{\partial q^{i}} +
\frac{\partial a_{ji}}{\partial q^{s}} = 0,
\eqno(4.8)
$$
$$
\frac{\partial b_{i}}{\partial q^{j}} -
\frac{\partial b_{j}}{\partial q^{i}} =
\frac{\partial a_{ij}}{\partial t}.
$$
By setting $a_{ij} =- (e/c) \epsilon_{ijk} B^{k}$ and
$e E^{i} = \delta^{ij} b_{j}$, we obtain the Lorentz force law
(3.12) after reversing the direction of the velocity vector, 
such that ${\bf E}$ and ${\bf B}$ satisfy,
$$
{\bf \nabla} \cdot {\bf B} = 0,
\qquad
{\bf \nabla } \times {\bf E} = - \frac{1}{c} 
\frac{\partial {\bf B}}{\partial t}.
$$
These results are consistent with the development from the
Poisson bracket point of view. It is interesting to note that if
we make the opposite selection $b_{i} = \delta_{ij} B^{j}$
and $a_{ij} = \epsilon_{ijk} E^{k}$, albeit for which there is no
known force law, the remaining two sourceless Maxwell equations appear.
Of course, the Maxwell equations are known to be symmetric under
the transformations ${\bf E} \rightarrow {\bf B}$ and
${\bf B} \rightarrow - {\bf E}$.

To determine the quantity ${\cal C}$ in (4.6), we have to be allowed
to reason from a physical point of view. This functional will
be a scalar formed from the fields ${\bf E}$ and ${\bf B}$,
which will serve as a Lagrangian for these fields. To this end, we
construct a functional of the form
$$
L_{em} = \int d^{3} x \, ( \alpha \, {\bf E}^{2} ({\bf x},t) + \beta 
\, {\bf B}^{2} ({\bf x},t)
+ \gamma \, {\bf E} ({\bf x},t) \cdot {\bf B}({\bf x},t)),
\eqno(4.9)
$$
and identify $L_{em}$ with the integration constant ${\cal C}$.
This reflects the fact that the fields involved will have an
energy density associated with them in the absence of particles.
Adding terms such as ${\bf A}^2$ and $A_{0}^2$ to this integrand
would destroy the form invariance of the total Lagrangian under
the addition of a total time derivative. 

Since ${\bf E}$ is a
vector and ${\bf B}$ is a pseudovector, this will be invariant
with respect to parity only if $\gamma=0$. The constants
$\alpha$ and $\beta$ can be taken so that other equations and
quantities obtained from $L$ will have their standard forms,
in this case, $\alpha = \beta=1/2$. Now, from the Lagrangian and
the principle of least action, we can immediately determine the remaining
two Maxwell equations. If the action is varied with respect to $A_{0}$,
Gauss's law results and if we vary with respect to ${\bf A}$,
then the Amp\'ere-Maxwell law results as before.

To summarize, it has been shown that a set of basic Poisson
brackets leads to some of the basic structures in electromagnetism,
in particular, the Lorentz force law. This can then be used in turn
to develop a Lagrangian which must exist on account of the
Helmholtz conditions. It is also quite interesting
that without any direct appeal to Lorentz invariance, 
a system of equations results which can be shown to be
relativistically invariant under Lorentz transformations {\bf [11-12]}.

It can be seen that the Maxwell equations are partitioned into two
groups, and this is clearly indicated in the details
of the reconstruction. Both Faraday's law (2.21) and (2.22)
are kinematical laws.
They follow from the relationship between the actual fields
and the charged particles. It is not surprising then that
these two equations are generated by the set of Poisson
brackets {\bf [13]}. On the other hand, the Amp\'ere-Maxwell law (2.18) and
Gauss's law (2.20) are dynamical equations. It would be of interest
to know whether other types of equations of physical interest,
for example to the realm of gravity {\bf [14]},
can be developed along similar lines as described here.

\begin{center}
{\bf References.}
\end{center}

\noindent
$[1]$ F. Dyson, Am. J. Phys. {\bf 58}, 209 (1990).   \\
$[2]$ P. Bracken, Int. J. Theor. Phys. {\bf 35}, 2125 (1996).      \\
$[3]$ P. Bracken, Int. J. Theor. Phys. {\bf 37}, 1625 (1998).    \\
$[4]$ D. Kobe, Am. J. Phys. {\bf 49}, 581, (1981).  \\
$[5]$ D. Kobe and E. C-T. Wen, J. Phys. A: Math. Gen. {\bf 15}, 787, (1982).  \\
$[6]$ P. Bracken, Acta. Appl. Math. {\bf 57}, 83, (1999).       \\
$[7]$ D. Kobe and E. C-T. Wen, J. Phys. A: Math. Gen. {\bf 13}, 3171 (1980).  \\
$[8]$ K. H. Yang, Ann. Phys. NY, {\bf 101}, 62, 1976.   \\
$[9]$ H. Goldstein, Classical Mechanics, Addison-Wesley, Reading,
Massachusetts (1950).   \\
$[10]$ P. Bracken, Tensor, {\bf 63}, 51, (2002).  \\
$[11]$ A. Berard and H. Mohrbach, Int. J. Theor. Phys.
{\bf 39}, 2623 (2000).   \\
$[12]$ A. Berard, Y. Grandati and H. Mohrbach, J. Math. Phys. {\bf 40}, 3732 (1999).  \\
$[13]$ J. W. van Holten, Nuclear Physics {\bf B 356}, 3, (1991).   \\
$[14]$ S. Tanimura, Annals of Physics, {\bf 220}, 229, (1992).  \\

\end{document}